\documentclass[aps,twocolumn,showpacs,preprintnumbers,amsmath,amssymb]{revtex4-1}

\usepackage{epsfig}
\usepackage{graphicx}

\usepackage{amsmath}
\usepackage{amssymb}
\usepackage{accents}
\usepackage{color}        %{\color{red/green/blue}........text.........}
\usepackage{slashed}

\newcommand{\bw}{\begin{widetext}}
\newcommand{\ew}{\end{widetext}}
\newcommand{\be}{\begin{equation}}
\newcommand{\ee}{\end{equation}}
\newcommand{\bea}{\begin{eqnarray}}
\newcommand{\eea}{\end{eqnarray}}
\newcommand{\mh}{\mathcal{H}}
\newcommand{\mr}{\mathcal{R}}
\newcommand{\lsp}{\left(}
\newcommand{\rsp}{\right)}
\newcommand{\lsb}{\left[}
\newcommand{\rsb}{\right]}
\newcommand{\lsc}{\left\{}
\newcommand{\rsc}{\right\}}
\newcommand{\nn}{\nonumber}
\newcommand{\vp}{\varphi}
\newcommand{\ba}{\begin{appendix}}
\newcommand{\ea}{\end{appendix}}
\begin{document}

\title{A note on inflationary scenario with non-minimal coupling}

\author{Abhishek Basak and Jitesh R. Bhatt}
%\emailAdd{abhishek@prl.res.in}
%\author{Jitesh R. Bhatt}
%\emailAdd{jeet@prl.res.in}
\affiliation{Physical Research Laboratory, Navarangpura, Ahmedabad, India}

\date{\today} %%
%\revised{}
% \accepted{} %% These are for published papers.
%\preprint{}% OR: \preprint{Aaaa/Mm/Yy\\Aaa-aa/Nnnnnn}
% Use \hepth etc. also in bibliography. 

\begin{abstract}
%In the present manuscript 
%We consider non-minimally coupled scalar-field inflaton and calculate
%the first order perturbation theory without assuming an adiabatic equation
%of state.

We consider first order perturbation theory for a non-minimally coupled
inflaton field without assuming an adiabatic equation of state. 
In general perturbations in non-minimally coupled theory may be
non-adiabatic. However under the slow-roll assumptions the perturbation
theory may look like adiabatic one. We show in the frame-work of
perturbation theory, that our results of spectral index and bound on
no-minimal coupling parameter agree with the results obtained
using the adiabatic equation of state by the earlier authors. 
%by the
%earlier authors with the adiabatic equation of state.
\end{abstract}

%  \abstract{We study the first order perturbations in a non-minimally
%  coupled scalar field theory in the Jordan frame.
% It is shown by comparing the  expression for the spectral index $n_{\mr}$
% with the observed value that the non-minimal coupling constant $\xi$ can have an upper bound.
% The bound on $\xi$ is generic as it is obtained without assuming any specific form of potential.
% Source of this bound is due to the presence of 
%  $\beta=\frac{\dot f}{Hf}$ term in the Friedman equation, where $f$ signifies the non-minimal coupling
% term.} 

%\begin{document}
\maketitle

\section{Introduction:}
Inflationary paradigm has become extremely useful in solving many 
problems with the standard big-bang theory and very successful in predicting
the fluctuations in the observed cosmic microwave background
radiation\cite{Guth:1980zm,Riotto:2002yw}. However, 
the nature of the inflaton potential remains to be
uncertain and as a consequence there has been a variety of models  through which
the inflationary paradigm can be implemented\cite{Linde:2007fr} . One of the important class of such models that
has been extensively investigated in recent time is that of a
non-minimally coupled scalar
fields\cite{Futamase:1989cn,Komatsu:1999mt,Sakai:1998rg,Faraoni:1996rf,Makino:1991dc,Komatsu:1998ju,CervantesCota:1994zf,Gupta:2009kk,Setare:2008mb}.
In most of the model of the inflation the mass of the scalar field is considered
to be around $10^{13}$~GeV and the extremely small value for the 
strength of the quartic self-coupling $\lambda \sim
10^{-13}$\cite{Bezrukov:2007ep}. This 
high value of the mass is considered to be an evidence for the physics beyond
the standard model. In Refs.\cite{Fakir:1990ab} it was shown that in case of
chaotic inflation model instead of minimally coupled
theory if non-minimal coupling is taken then amplitude of
density perturbation will constrain the ratio $\frac{\lambda}{\xi^{2}}$ rather
than $\lambda$ 
%($\xi$ is the non-minimal coupling and $\lambda$ is the self-coupling parameter.) 
Therefore in minimally coupled theory one can remove
the tight constraint on the 
self coupling parameter $\lambda$ by choosing a higher value of $\xi$. It
is customary to choose the value of $\xi\sim10^{3}$.
However, an extremely interesting possibility of having a non-minimally coupled
standard model Higgs field as an inflaton has been pointed out in Ref.\cite{Bezrukov:2007ep}. 
In this model for a sufficiently large strength of the non-minimal coupling
$1\ll\sqrt{\xi}\lll 10^{17}$, 
it is possible to have $\lambda \approx 1$.
Inert Higgs doublet has also been studied
which gives the scale invariant density perturbation\cite{Das:2011wm}.
It is well known  \cite{Dicke:1962ab} 
that the non-minimally coupled theory(Jordan frame) can be transformed into a minimally coupled
theory(Einstein frame) by a set of transformations of the metric and the
field. It should be noted that a consistent formalism of first order
perturbation theory \cite{Hwang:1990ac,Hwang:1990ab,
PhysRevD.54.1460,Hwang:1998ab,Sugiyama:2010ab} and various bound on $\xi$
\cite{Kaiser:1995nv,Kaiser:1994vs,Fakir:1990ab} has been studied  by several earlier authors.
Further, we would like to say in
Refs.\cite{Hwang:1990ac,Hwang:1990ab, PhysRevD.54.1460,Kaiser:1995nv,Hwang:1998ab}
a `gauge ready' approach was developed to study the first order perturbation theory. 
In this approach the variable under a
gauge condition which removes the gauge mode completely can be considered
as the gauge invariant one. It has been shown recently in Ref.\cite{Brown:2011eh} that
all other gauges except Newtonian gauge is being tangled in the
transformation between these two frames.

\par
In this work we are studying the first order perturbation in the the
context of 
non-minimally coupled inflationary theory. We would like to emphasize that
in order to derive 
equation of motion for first order curvature perturbation in the previous
works an adiabatic equation of state was used. But in general the adiabatic
equation of state for non-minimally coupled theory may not be possible
\cite{PhysRevD.54.1460,Carloni:2006gy}. Recently the non-adiabatic evolution of curvature
perturbation is shown to exist in the case of non-minimally coupled
multi-field inflation scenario \cite{White:2012ab}. 
Therefore, we believe that it is important to relax the assumption about the
equation of state for a single field inflaton theory. In this work we calculate
the first order curvature perturbation and the spectral index without assuming
any equation of state. Our results are consistent with the results obtained by earlier workers.

%Therefore, in the
%present work we calculate first order curvature perturbation and the
%spectral index without assuming anything about the equation of state. Our
%results are consistent with the results obtained by earlier workers.

As mentioned earlier , for the case of chaotic inflation in Jordan frame
that the density perturbation on Jordan frame are constrained by the ratio
$\frac{\lambda}{\xi^{2}}$ and this may allow for the values $\xi\gg1$.
Further it is to be noted that in Ref. \cite{Kaiser:1994vs} in order to
produce the Harrison-Zel'dovich spectrum ($n_{\mr}\sim1$) for the
`new-inflation' scenario it was necessary to put an upper bound on $\xi$.
However no such bound was required for the chaotic inflation scenario. I
this work we show that in general there exist two branches, in one branch
$\dot\vp$ (inflaton velocity) is negative and $\dot\vp$ is positive in the
other branch depending on the nature of the potential. For those classes of
potential where $\dot\vp>0$ one must have an upper bound on $\xi$. But the other
branch may allow a large value of $\xi$ for some classes of the potentials
provided the slow-roll parameter $\epsilon_{V}$ is very small. It
seems that for power-law kind of potentials only $\lambda\vp^{4}$ can allow
large value of $\xi$ which is consistent with the already known results.

The action in Jordan frame is given by
\begin{equation}
S=\int{d^{4}x\sqrt{-g}\left[f\left(\varphi\right)R-\frac{1}{2}\varphi_{;\mu}\varphi^{;\mu}+V\left(\varphi\right)\right]},
\label{eq:actionj}
\end{equation}
in case of non-minimal coupling $f\lsp\vp\rsp=1+\xi\vp^{2}$, where $\vp$ is
the scalar field, $\xi$ is a constant, $R$ is the Ricci scalar and $V(\vp)$
is the potential. Here we have considered $M_{pl}=1$. The field equation in
Jordan frame is given as 
\bea
\label{eq:fe}
\nn
f\left(\varphi\right)&&\left(R_{\mu\nu}-\frac{1}{2}g_{\mu\nu}R\right)=\frac{1}{2}\left(\varphi_{;\mu}\varphi_{;\nu}-\frac{1}{2}g_{\mu\nu}\varphi_{;\alpha}\varphi^{;\alpha}\right)+\\
&&\frac{1}{2}g_{\mu\nu}V\left(\varphi\right)+f\left(\varphi\right)_{;\mu;\nu}-g_{\mu\nu}\Box f\left(\varphi\right)
\eea
Jordan frame action (\ref{eq:actionj}) can be transformed into the Einstein frame action
\be
S=\int{d^{4}x\sqrt{-\hat g}\left[\frac{1}{2}\hat
R-\frac{1}{2}\sigma_{;\mu}\sigma^{;\mu}+\hat V\left(\sigma\right)\right]},
\label{eq:actione}
\ee
by transforming the metric and the scalar field in the following way,
\be
\hat g_{\mu\nu}=2f\lsp\vp\rsp g_{\mu\nu}, \quad
\lsp\frac{d\sigma}{d\vp}\rsp^{2}\equiv\frac{1}{2f}\lsp1+3\frac{f^{2}_{,\vp}}{f}\rsp.
\label{eq:trans}
\ee
where $\hat\quad$ represents the Einstein frame and $\hat
V(\sigma)=\frac{V(\vp)}{4f^{2}}$. Using \ref{eq:trans} the field
equation in Eq.(\ref{eq:fe}) can be shown to be transformed into the Einstein equations.
This is true for the perturbed field equations also. In Einstein frame the
equation of motion for the comoving curvature perturbation
$\hat{\mr}=\hat\psi+\hat{\mh}\frac{\delta\sigma}{\sigma^{\prime}}$ can be written as 
\be
\hat\mr^{\prime\prime}+\lsb\ln\hat\gamma\rsb^{\prime}\hat\mr^{\prime}+k^{2}\hat\mr=0
\label{eq:eomre}
\ee
 where $\hat\gamma=\frac{\hat a^{2}\sigma^{\prime 2}}{\hat{\mh}^{2}}$,
 $\hat{\psi}$ and $\hat{\mh}$ are metric perturbations and Hubble parameter
 in Einstein frame. In
 Einstein frame spectral index $\hat n_{\mr}$ can be expressed in terms of
 slow roll parameters as 
 \be
 \hat n_{\mr}-1=-4\hat\epsilon_{V}-2\hat\delta.
 \label{eq:sie}
 \ee
 where $\hat\epsilon_{V}=\frac{3}{2}\frac{\hat{\sigma}^{\prime
 2}}{\hat a^{2}\hat V}$ and $\hat\delta=1-\frac{\hat\sigma^{\prime\prime}}{\hat\mh\hat\sigma^{\prime}}$ are the slow roll
 parameters in the Einstein frame.

%%%%%%%%%%%%%%%%%%%%%%%%%%%%%%%%%%%%%%%%%%%%%%%%%%%%%%%%%%%%%%%%%%%%%%%%%%%%%%%%%%%%%%%%%%%%%%%%%%%%%%%%%%%%%%%%%%%%%%%%%%%%%%%%%%%%%%%%%%%%%%%%%%%%%%%%%%%%%%

\section{Slow roll parameters in Jordan Frame:} Before we
calculate the equation of motion 
of $\mr$ we define the slow-roll parameters in the Jordan
frame. Equation for the scalar field $\vp$ in Jordan frame is given by
\begin{equation}
\label{eq:eom}
\varphi^{\prime\prime}+2\mathcal{H}\varphi^{\prime}+a^{2}f_{\varphi}R+a^{2}V_{\varphi}=0,
\end{equation}
where
$R=-\frac{6}{a^{2}}\left(\mathcal{H}^{\prime}+\mathcal{H}^{2}\right)$. Here
$\prime$ denotes the derivative with respect to conformal time $\eta$, this
is related to natural time as $d\eta=\frac{dt}{a}$. $\mh$ is the Hubble
parameter in conformal time defined as $\mh=\frac{a^{\prime}}{a}$ and $a$
is the scale factor. Friedman equations in Jordan frame gives us the
following relations:
\bea
\mh^{2}\lsp1+\beta\rsp&=&\frac{1}{6f}\lsp\frac{\vp^{\prime
2}}{2}+a^{2}V\rsp, \label{eq:fra0}\\ 
\mh{^2}-\mh^{\prime}&=&\frac{\vp^{\prime
2}}{4f}+\frac{f^{\prime\prime}}{2f}-\mh\frac{f^{\prime}}{f}.\label{eq:fra}
\eea
In the case of the standard inflation we may say
that the inflaton is slowly rolling down the potential, therefore
the assumptions of slow roll are 
% $\dot\vp^{2}\ll V$ and
% $\ddot\vp\ll H\dot\vp$. Here $dot$ denotes the derivative with respect to
% the cosmic time $t$. We define the slow roll parameters $\epsilon_{V}$ and
%$\delta$ as 
\be
\epsilon_{V}=\frac{3}{2}\frac{\vp^{\prime 2}}{a^{2}V}\ll1, \qquad
\delta=1-\frac{\vp^{\prime\prime}}{\mh\vp^{\prime}}\ll1.
\label{sr1}
\ee
In case of inflation driven by the scalar field minimally coupled to
gravity $-\frac{\dot H}{H^{2}}=1-\frac{\mh^{\prime}}{\mh^{2}}=\epsilon$, therefore $\epsilon\ll1$ implies
the smallness of $\dot H$ compared to $H^{2}$, i.e, $H$ remains almost
constant during inflation. 
But for the inflationary scenario in the Jordan frame 
smallness of $\epsilon$ alone does not ensure that $H$ will remain constant during inflation. 
%But in case of the present inflationary theory
%where
%inflaton is coupled to gravity smallness of $\epsilon$ alone does not ensure
%that $H$ will remain constant during inflation. 
% Eq.(\ref{eq:fra}) gives us,
% \bea
% \nn-\frac{\dot H}{H^{2}}&=&\lsp1-\frac{\mh^{\prime}}{\mh^{2}}\rsp\\
%                      &=&\frac{\vp^{\prime
%                      2}}{4f\mh^{2}}+\frac{f^{\prime\prime}}{2f\mh^{2}}-\frac{f^{\prime}}{f\mh}.
% \label{eq:fra2}                     
% \eea
% If we define two parameters $\lambda\equiv\frac{f^{\prime}}{f\mh}$ and
% $\alpha\equiv\frac{f^{\prime\prime}}{\mh f^{\prime}}$, we can write
% Eq.(\ref{eq:fra2}) as 
% \be
% -\frac{\dot H}{H^{2}}=\lsp1-\frac{\mh^{\prime}}{\mh^{2}}\rsp=\frac{\vp^{\prime
% 2}}{4f\mh^{2}}+\frac{\alpha\lambda}{2}-\lambda.
% \label{eq:fra3}
% \ee
Finally using Eq. (\ref{eq:fra0}) we can write Eq. (\ref{eq:fra}) as:
\be
\lsp1-\frac{\mh^{\prime}}{\mh^{2}}\rsp=\epsilon_{V}\lsp1+\beta\rsp+\frac{\alpha\beta}{2}-\beta.
\label{eq:fra4}
\ee
Here we have defined $\beta\equiv\frac{f^{\prime}}{f\mh}$ and
$\alpha\equiv\frac{f^{\prime\prime}}{\mh f^{\prime}}$. At this point if we
consider $\epsilon_{V}\ll1$, $\alpha\ll1$ and $\beta\ll1$
and neglect the second and third term on the right hand side of the above
expression, still we have 
\be
-\frac{\dot H}{H^{2}}=\lsp1-\frac{\mh^{\prime}}{\mh^{2}}\rsp\simeq\epsilon_{V}-\beta.
\label{eq:fra5}
\ee
So, we can see that smallness of $\beta$ and $\alpha$ are required to
drive inflation.
Therefore the end of inflation is not determined by $\epsilon_{V}$ only.
Inflation ends when $\epsilon_{V}-\beta\sim1$.

%%%%%%%%%%%%%%%%%%%%%%%%%%%%%%%%%%%%%%%%%%%%%%%%%%%%%%%%%%%%%%%%%%%%%%%%%%%%%%%%%%%%%%%%%%%%%%%%%%%%%%%%%%%%%%%%%%%%%%%%%%%%%%%%%%%%%%%%%%%%%%%%%%%%%%%%%%%%%%
\section{Perturbed equations:}
In this work we are writing the total metric (background and perturbed) in
Newnotian gauge as,
\be
g_{\mu\nu}= g^{(0)}_{\mu\nu}+\delta g_{\mu\nu}=a^{2} \begin{pmatrix}
                 \lsp1+2\phi\rsp & \mathbb{O}\\
                 \mathbb{O} & \lsc\lsp-1+2\psi\rsp\delta_{ij}\rsc
                 \end{pmatrix}.
\ee
It is useful to note that 
metric perturbations, and $\mh$ follow the transformations 
\be
\hat\psi=\psi-\frac{f_{1}}{2f}, \qquad  \hat\phi=\phi+\frac{f_{1}}{2f},
\qquad  \hat\mh=\mh+\frac{f^{\prime}}{2f},
\label{eq:transp}
\ee
when we go to Einstein frame from Jordan frame, where $f_{1}=\delta f=f_{\vp}\delta\vp$.
Perturbing the Eq.(\ref{eq:fe}) we get the following equations. Time-space
component of field equation is:
\begin{equation}
\label{eq:pts}
\psi^{\prime}+\mathcal{H}\phi=\frac{\varphi^{\prime
2}}{4\mathcal{H}f}\left(\mathcal{H}\frac{\delta\varphi}{\varphi^{\prime}}\right)
-\frac{\mathcal{H}f_{1}}{2f}-\phi\frac{f^{\prime}}{2f}+\frac{f_{1}^{\prime}}{2f}.
\end{equation}
Off diagonal space-space component is, 
\begin{equation}
\label{eq:poss}
\psi-\phi=\frac{f_{1}}{f}.
\end{equation}
% Using Eq.(\ref{eq:poss}) one can write Eq.(\ref{eq:pts}) as 
% \bea
% \label{eq:ptsa}
% \nn\psi^{\prime}+\mathcal{H}\phi&=&\frac{\varphi^{\prime
% 2}}{4\mathcal{H}f}\left(\mathcal{H}\frac{\delta\varphi}{\varphi^{\prime}}\right)
% -\frac{\mathcal{H}f_{1}}{2f}-\lsp\psi-\frac{f_{1}}{f}\rsp\frac{f^{\prime}}{2f}+\\
% &&\frac{f_{1}^{\prime}}{2f}
% \eea
Time-time component of the field equation is,
\bea
\label{eq:ptt}
\nn\left(3\mathcal{H}^{2}\right)f_{1}&&+2\left[\Delta\psi-3\mathcal{H}\left(\psi^{\prime}+\mathcal{H}\phi\right)\right]f=\frac{1}{2}\varphi^{\prime}\delta\varphi^{\prime}+\Delta\left(f_{1}\right)-\\
&&3\mathcal{H}f_{1}^{\prime}+3\psi^{\prime}f^{\prime}+\frac{1}{2}a^{2}V_{\varphi}\delta\varphi+6\mathcal{H}\phi
f^{\prime}-\frac{1}{2}\phi\varphi^{\prime 2}.
\eea
The above equations are written in Jordan frame.
Using using the transformations 
defined by equation (\ref{eq:trans}) along with  (\ref{eq:transp}),
equations  
(\ref{eq:pts}), (\ref{eq:poss}) and (\ref{eq:ptt}) transform into the corresponding equations
in the Einstein frame considered in the literature, for example Ref.\cite{Mukhanov:2005ab}.
%Keeping in mind that derivative with respect to conformal time does
%not change when we change the frames, one can check that equations
%(\ref{eq:pts}), (\ref{eq:poss}) and (\ref{eq:ptt}) remains invariant,
%i.e., when we apply the transformations (\ref{eq:trans}) along with (\ref{eq:transp}) in the Jordan frame, we get the perturbed equations in
%the Einstein frame given in inflation literatures, for example
%Ref.\cite{Mukhanov:2005ab}.
Next, we  write down some useful relations which  will be used frequently
in later discussion. From the definition of $f_{1}$ given above in terms of $\beta$ we can
write, 
\bea
\nn\frac{f_{1}}{f}&=&\beta\lsp\mh\frac{\delta\vp}{\vp^{\prime}}\rsp\\
\frac{f_{1}^{\prime}}{f}&\simeq&\mh\beta\lsp\frac{\gamma}{\mh}-1\rsp\mh\frac{\delta\vp}{\vp^{\prime}}+\beta\lsp\mh\frac{\delta\vp}{\vp^{\prime}}\rsp^{\prime},
\label{eq:usfr1}
\eea
where we define $\gamma\equiv\frac{\vp^{\prime 2}}{4f\mh}$. Note that $\gamma$
is not a slow roll parameter. Here we have dropped the terms proportional
to second order in slow roll parameters $\beta$ and $\alpha$.
Similarly (\ref{eq:fra}) can be written as 
\be
\mh^{2}-\mh^{\prime}\simeq\mh\gamma-\mh^{2}\beta.
\label{eq:fra1}
\ee
At this point let us define two variables $Y$ and $\mr$ as following 
\be
Y=\frac{a^{2}}{\mh}\psi, \qquad \mr={\psi+\mh\frac{\delta\vp}{\vp^{\prime}}}.
\label{eq:nv}
\ee
It should be noted that $\hat\mr$ and $\mr$ are invariant under the
transformations \ref{eq:trans}.
Substituting the expression of $\phi$ in (\ref{eq:poss}), in linear order
of $\beta$ we can write
(\ref{eq:pts}) in terms of $Y$ and $\mr$ as,
\be
Y^{\prime}\simeq\frac{a^{2}\gamma}{\mh}\mr-\frac{a^{2}\beta}{2}\lsp2\frac{\mh}{a^{2}}Y-\frac{\mr^{\prime}}{\mh}\rsp,
\label{eq:ptsa3}
\ee
where we have used (\ref{eq:usfr1}) and (\ref{eq:fra1}).
\par
Using the background equation of motion of $\vp$ we can eliminate $V_{\vp}$ from
(\ref{eq:ptt}) and finally using (\ref{eq:poss}), (\ref{eq:usfr1}) and
(\ref{eq:fra1}) we can write (\ref{eq:ptt}) as 
\be
\Delta\lsp\frac{\mh}{a^{2}}Y\rsp\simeq\gamma\lsp1-\beta\rsp\mr^{\prime}-\frac{\beta\gamma}{2}\lsp\gamma\mr-\frac{\mh}{a^{2}}Y^{\prime}\rsp+\frac{\beta}{2}\lsp\Delta\mr\rsp.
\label{eq:ptta6}
\ee
\par
Using the above equations one can calculate the power spectrum in the usual
way. 
Equations (\ref{eq:ptsa3}) and (\ref{eq:ptta6}) are coupled equations.
Decoupling these two equations would lead to the follwing equation for
$\mr$: 
\bea
\nn\mr^{\prime\prime}&+&\lsb\lsc\ln\lsp\frac{a^{2}}{\mh}\gamma\lsp1-\beta\rsp\rsp\rsc^{\prime}+\mh\beta\rsb\mr^{\prime}+\\
&&k^{2}\lsb1+\beta-\lsp\frac{\beta a^{2}}{2\mh}\rsp^{\prime}\frac{\mh}{a^{2}\gamma}\rsb\mr=0.
\label{eq:eomr}
\eea
As $\prime$, $k^{2}$, $\mr$ remains invariant under the transformation of the
frames, we can see that Eq.(\ref{eq:eomre}) and Eq.(\ref{eq:eomr}) does not
transform to each other when we consider the transformations listed above.
One can identify the coefficient of $k^{2}$ as the square of sound speed of
perturbation namely $C_{s}^{2}$,
\be
C_{s}^{2}\simeq1+\beta-\lsp\frac{\beta a^{2}}{2\mh}\rsp^{\prime}\frac{\mh}{a^{2}\gamma}.
\label{eq:cs}
\ee
It is possible to  write the last term in the above expression in terms of slow roll
parameters and one can write the expression for $C_{s}^{2}$ as 
\be
C_{s}^{2}=1+\frac{\beta}{2}-\frac{\beta}{2\epsilon_{V}}.
\label{eq:cs1}
\ee
Here it should be noted that the value of $\frac{\beta}{2\epsilon_{V}}$
can't exceed $\frac{1}{2}$ as $\beta<\epsilon_{V}$ is required to be
satisfied when $\beta$ is positive (see section (\ref{s5})).
\par

%%%%%%%%%%%%%%%%%%%%%%%%%%%%%%%%%%%%%%%%%%%%%%%%%%%%%%%%%%%%%%%%%%%%%%%%%%%%%
\section{Power spectrum and spectral index:}
To calculate the power spectrum and spectral index we follow the standard
procedure given by Mukhanov\cite{Mukhanov:2005ab}. We first substitute
$\mr=m.v$ into Eq.(\ref{eq:eomr}), where $m$ is a function of $\eta$ only,
i.e., $m=m\lsp\eta\rsp$ and
$v=v\lsp\eta,\vec{k}\rsp$. 
% Then the equation of $v$ becomes 
% \be
% v^{\prime\prime}+\lsp2\frac{m^{\prime}}{m}+A\rsp v^{\prime}+\lsp
% A\frac{m^{\prime}}{m}+\frac{m^{\prime\prime}}{m}+B\rsp v=0,
% \label{eq:eomv}
% \ee
% where
% $A=\lsb\frac{\lsc\frac{a^{2}}{\mh}\gamma\lsp1-\lambda\rsp\rsc^{\prime}}{\lsc\frac{a^{2}}{\mh}\gamma\lsp1-\lambda\rsp\rsc}+\mh\lambda\rsb$
% and $B=C_{s}^{2}k^{2}=k^{2}\lsb1+\lambda-\lsp\frac{\lambda
% a^{2}}{2\mh}\rsp^{\prime}\frac{\mh}{a^{2}\gamma}\rsb$. Now we choose $m$
% such that the coefficient of $v^{\prime}$ becomes zero. Therefore we get 
% \be
% \frac{m^{\prime}}{m}=-\frac{A}{2}.
% \label{eq:m}
% \ee
% Substituting this expression of $m$ in Eq.(\ref{eq:eomv}) we find the equation
% for $v$ as 
Removing the $v^{\prime}$ term by setting the coefficient of $v^{\prime}$
to be zero, which yields $\frac{m^{\prime}}{m}=-\frac{A}{2}$, we get the
equation of $v$
\be
v^{\prime\prime}+\lsb C_{s}^{2}k^{2}-\frac{A^{\prime}}{2}-\frac{A^{2}}{4}\rsb v=0.
\label{eq:eomv1}
\ee
Here
$A=\lsb\frac{\lsc\frac{a^{2}}{\mh}\gamma\lsp1-\beta\rsp\rsc^{\prime}}{\lsc\frac{a^{2}}{\mh}\gamma\lsp1-\beta\rsp\rsc}+\mh\beta\rsb$.
Now this equation can be written as Bessel differential equation and can be solved exactly in terms of Hankel functions once
we write $A^{\prime}$ and $A^{2}$ as $\frac{1}{\eta^{2}}$. $A$
and $A^{\prime}$ can be expressed 
in terms of slow roll parameters as
\bea
\nn A&=&2aH\lsp1+\epsilon_{V}-\beta+\delta\rsp,\\
A^{\prime}&=&2a^{2}H^{2}\lsp1-\epsilon_{V}+\beta\rsp\lsp1+\epsilon_{V}-\beta+\delta\rsp.
\label{eq:A}
\eea
Here we have used (\ref{eq:fra5}).
From
(\ref{eq:fra5}) we can also notice that in this present case we can write
$aH\simeq-\frac{1}{\eta}\frac{1}{1-\epsilon_{V}+\beta}$. Using this
expression we can write
$\frac{A^{\prime}}{2}+\frac{A^{2}}{4}=\frac{1}{\eta^{2}}\lsp\nu^{2}-\frac{1}{4}\rsp$,
where
$\nu=\frac{1+\epsilon_{V}-\beta+\delta}{1-\epsilon_{V}+\beta}+\frac{1}{2}\simeq\lsp\frac{3}{2}+2\epsilon_{V}-2\beta+\delta\rsp$.
Therefore we finally write Eq.(\ref{eq:eomv1}) as
\be
v^{\prime\prime}+\lsb
C_{s}^{2}k^{2}-\frac{1}{\eta^{2}}\lsp\nu^{2}-\frac{1}{4}\rsp\rsb v=0.
\label{eq:eomv2}
\ee
Considering the fact that $C_{s}$ is a constant we can express the solution
of this equation in terms of the Hankel functions,
\be
v_{k}=\lsp-\eta\rsp^{1/2}\lsb
A_{1}H_{\nu}^{\lsp1\rsp}\lsp-C_{s}k\eta\rsp+A_{2}H_{\nu}^{\lsp2\rsp}\lsp-C_{s}k\eta\rsp\rsb
\label{eq:modfn}
\ee
Matching this solution of $v$ with the free quantum field for
$\frac{C_{s}k}{aH}\gg1$(short wave length limit), we choose 
\be
A_{1}=\frac{\sqrt{\pi}}{2}exp\lsb
i\frac{\pi}{2}\lsp\nu+\frac{1}{2}\rsp\rsb, \qquad
A_{2}=0.
\ee
For long wavelength ($\frac{C_{s}k}{aH}\ll1$) we find that $v_{k}$ behaves
as
\be
v_{k}\propto k^{-\nu}.
\ee
Next writing this solutions in terms of $\mr$ we get the power spectrum
as 
\be
\mathcal{P}_{\mr}\propto k^{3-2\nu}.
\ee
Spectral index is defined as 
\be
n_{\mr}-1=\frac{d \ln\mathcal{P}_{\mr}}{d \ln k}.
\ee
Therefore we get the expression of spectral index as 
\be
n_{\mr}-1=3-2\nu.
\label{eq:sij}
\ee
Substituting the expression of $\nu$ in the above equation we get the
spectral index as 
\be
n_{\mr}=1-4\epsilon_{V}+4\beta-2\delta.
\label{eq:sij1}
\ee
In this expression we get an additional term $4\beta$ along with the
standard terms we get from the minimal coupling case.

%%%%%%%%%%%%%%%%%%%%%%%%%%%%%%%%%%%%%%%%%%%%%%%%%%%%%%%%%%%%%%%%%%%%%%%%%%%%%%%%

%%%%%%%%%%%%%%%%%%%%%%%%%%%%%%%%%%%%%%%%%%%%%%%%%%%%%%%%%%%%%%%%%%%%%%%%%%%%%%%%

%%%%%%%%%%%%%%%%%%%%%%%%%%%%%%%%%%%%%%%%%%%%%%%%%%%%%%%%%%%%%%%%%%%%%%%%%%%%%%%%%%%%%%%%%%%%%%%%%%%%%%%%%%%%%%%%%%%%%%%%%%%%%%%%%%%%%%%%%%%%%%%%%%%%%%%%%%%%%%%%
\section{Various values of $\xi$:\label{s5}}

\subsection{Spectral index:\label{ss1}}
In the appendix the first order perturbation theory is done and expression
of spectal index is given (\ref{eq:sij1}) in Jordan frame. In this
calculation we have not 
assumed any specific equation of state unlike the previous authors
\cite{Hwang:1990ac,Hwang:1990ab}. It is shown later that expression
($n_{\mr}=1-4\epsilon_{V}+4\beta-2\delta$)
obtained here matches exactly with Ref.\cite{Hwang:1990ac,Hwang:1990ab} in
the region $\xi\vp^{2}\gg1$. WMAP 7 years data suggests
that $n_{\mr}=0.968\pm0.012$ \cite{Komatsu:2010fb}. In this case let us
consider that $\delta=0$. Therefore the expression of $n_{\mr}$ suggests
that $\epsilon_{V}-\beta\sim10^{-2}$. This condition is consistent
with the condition that $0<-\frac{\dot H}{H^{2}}\ll1$. The lower bound is
necessary to have accelerated expansion. The expression of $\beta$
can be written as:
\be
\beta=\frac{\dot
f}{Hf}=\frac{2\xi\vp}{\sqrt{f}}\frac{\dot\vp}{H\sqrt{f}}.
\label{lmd1}
\ee
Using the expression of $\epsilon_{V}$ (\ref{sr1}) and the Friedmann
equation (\ref{eq:fra0}) one can write the following identity:
\be
\frac{\dot{\vp}^{2}}{fH^{2}}\approx4\epsilon_{V}.
\label{lmd2}
\ee
Therefore for any potential in general case one can have two values of
$\beta$ in the region $\xi\vp^{2}\gg1$:
\be
\beta=\pm4\sqrt{\xi}\sqrt{\epsilon_{V}}.
\label{lmd3}
\ee
As $\beta$ can be written in terms of the scalar field $\vp$
($\beta=2\frac{\dot\vp}{H\vp}$), which sign to be picked will be
dependent entirely on the dynamics of the scalar field for any specific
potential.

Using the Friedmann equation and writing the equation of motion
(\ref{eq:eom}) in cosmic time one can find the expression (2.13) in
Ref.\cite{Komatsu:1998ju}:
\be
3H\dot\vp\approx\frac{1}{1+\xi\vp^{2}\lsp1+6\xi\rsp}\lsb4\xi\vp
V-\lsp1+\xi\vp^{2}\rsp V_{,\vp}\rsb
\label{eom1}
\ee
Therefore $\beta$ becomes negative if 
\be
4\xi\vp V-\lsp1+\xi\vp^{2}\rsp V_{,\vp}<0,
\label{con1}
\ee
$V=\frac{1}{4}\lambda\vp^{4}$ is an example of this branch.
Whereas $\beta$ becomes positive if 
\be
4\xi\vp V-\lsp1+\xi\vp^{2}\rsp V_{,\vp}>0,
\label{con2}
\ee
$V=\frac{1}{2}m^{2}\vp^{2}$ is an example for this branch.
When condition (\ref{con1}) is satisfied one can write
\be
-\frac{\dot H}{H^{2}}=\epsilon_{V}+4\sqrt{\xi}\sqrt{\epsilon_{V}}
\ee
In this case to have $-\frac{\dot H}{H^{2}}\sim10^{-2}$ one can choose a
large value of $\xi$, provided $\epsilon_{V}$ is very small. For example,
if we choose $\xi\sim10^{4}$ then $\epsilon_{V}\sim10^{-8}$. It turns out
that $\lambda\vp^{4}$ potential in the power-law class of potentials can give us
$\epsilon_{V}$ that small ($\sim10^{-8}$).    
    
In case of the
other branch with positive $\beta$ one can write the expression
(\ref{eq:fra5}) as
\be
-\frac{\dot H}{H^{2}}=\epsilon_{V}-4\sqrt{\xi}\sqrt{\epsilon_{V}}.
\ee
In this case to have $-\frac{\dot H}{H^{2}}>0$ along with $\beta\ll1$ we
have to have an additional condition on $\beta$: $\beta<\epsilon_{V}$. This
condition gives us as upper bound on the non-minimal coupling parameter
$\xi<\frac{\epsilon_{V}}{16}$. For a typical value of
$\epsilon_{V}\sim10^{-2}$ we have $\xi<10^{-3}$. In this branch one
can not make $\xi$ very large by choosing a smaller value $\epsilon_{V}$
like the previous case, as this may make $-\frac{\dot H}{H^{2}}<0$
violating the condition for accelerated expansion.

% Now let us consider a power-law potential $V=\lambda\vp^{n}$. In this kind
% of models of chaotic inflation expression (\ref{eom1}) can be written as 
% \be
% 3H\dot\vp\approx\frac{\lambda
% n\vp^{n-1}}{1+\xi\vp^{2}\lsp1+6\xi\rsp}\lsb\lsp\frac{4}{n}-1\rsp\xi\vp^{2}-1\rsb
% \label{eom2}
% \ee
% From the above expression it is evident that for $n\geq4$ the field
% velocity $\dot\vp<0$. Therefore one can have choose the value of $\xi$ very
% large. But this large value will make $\epsilon_{V}$ very small. Whereas in
% case of $n<4$ there are two possibilities, (i) if
% $\xi\vp^{2}>\frac{1}{\frac{4}{n}-1}$ then $\dot\vp>0$ and (ii)
% $\xi\vp^{2}<\frac{1}{\frac{4}{n}-1}$ then we have $\dot\vp<0$. But the case
% (ii) can be ignored as this case implies a weak coupling. Therefore in case
% of $n<4$ we will have a bound on $\xi$. 

%%%%%%%%%%%%%%%%%%%%%%%%%%%%%%%%%%%%%%%%%%%%%%%%%%%%%%%%%%%%%%%%%%%%%%%%%
\par
Further one may ask the question whether the 
formalism used here and the ``gauge-invariant" formalism presented in Refs.
\cite{PhysRevD.54.1460,Kaiser:1995nv} are equivalent? In other words one
can ask the question if the results obtained in this formalism is same as
the result obtained by previous authors. In what follows we address this
question:
The expression  of spectral index in those references is given by 
\be
n_{s}=1-4\epsilon-2\delta+2\beta-2\gamma, \label{spech}
\ee
where $\epsilon=-\frac{\dot H}{H^{2}}$, $\delta=\frac{\ddot\vp}{H\dot\vp}$,
$\gamma=\frac{\dot E}{2HE}$, $\beta=\frac{\dot f}{2Hf}$ and
$E=\lsp f+\frac{3}{2}f^{2}_{\vp}\rsp$. Using the expression of $\dot{H}$
and $H^{2}$ given in \cite{PhysRevD.54.1460} one can write
$\epsilon=\frac{\dot{\vp}^{2}}{2fH^{2}}+\frac{\ddot
f}{2fH^{2}}-\frac{\dot f}{2fH}$. The second term is a second order term and
can be ignored. Next, one can write $\epsilon\simeq\epsilon_{V}-\beta$ where
$\epsilon_{V}=\frac{\dot{\vp}^{2}}{2fH^{2}}$. So, (\ref{spech}) can be
written 
as $n_{s}=1-4\epsilon_{V}-2\delta+6\beta-2\gamma$. Next using the
definition of $E$ one can write 
\be
\gamma=\beta\frac{\lsp1+3f_{\vp\vp}\rsp
f}{f+\frac{3}{2}f^{2}_{\vp}}. 
\ee
Therefore one can see that $\gamma$ and
$\beta$ are not independent parameters. Using the definition of $f$ it can
be found that in the region we are interested in
$\xi\vp^{2}\gg1$ \cite{Bezrukov:2007ep}, $\gamma=\beta$. Therefore we find
$n_{s}=1-4\epsilon_{V}-2\delta+4\beta$. The parameter $\beta$ is similar to
the parameter $\beta$ we have used in this manuscript. Therefore we
get the similar kind of bound from earlier results also.

\subsection{Freezing out of curvature perturbation:\label{ss3}}
Now let us consider the evolution equation of $\mr$ in the super horizon
scale.
In $k\rightarrow0$ limit equation (\ref{eq:eomr}) can be written as: 
\be
\mr^{\prime\prime}+A\mr^{\prime}=0.\label{eq2}
\ee
where $A=2aH\lsp1+\epsilon_{V}-\beta+\delta\rsp.$
The general solution of the equation (\ref{eq2}) is:
\be
\mr=C_{1}+C_{2}\int\exp\lsb-\int Ad\eta\rsb d\eta, \label{eq3}
\ee
where $C_{1}$ and $C_{2}$ are constants of integration. 
Substituting the expression of
$A$ in equation (\ref{eq3}) one can rewrite the expression of $\mr$ as:

\begin{widetext}
\be
%\nn \mr&=&C_{1}+C_{2}\int\exp\lsb-2\int aH\lsp1+\epsilon_{V}-\beta+\delta\rsp d\eta\rsb d\eta,\\ 
\mr=C_{1}+C_{2}\int\lsb\exp\lsp2\int aH\beta
d\eta\rsp\times\exp\lsp-2\int aH\lsp1+\epsilon_{V}+\delta\rsp d\eta\rsp\rsb
d\eta.
\label{eq4}
\ee
\end{widetext}
The last exponential term is the standard term that comes in the minimally
coupled theory and the first exponential term is the contribution of the
non-minimal coupling. In general three cases can arise: (i) when $\beta=0$
the last term decays and one 
gets  $\mr=Constant$. This case is nothing but the minimal coupling
scenario. (ii) In case of negative value of $\beta$, the first term
again exponentially decays to give us a constant $\mr$ in the growing mode.
However (iii) if the parameter $\beta$ is positive then it is required that 
$\beta\ll1$ to stop the first exponential term from evolving and
one gets $\mr=Constant$ here also. From the time-time component of perturbed
field equation one gets at $k\rightarrow0$ limit:
\be
\mr^{\prime}\simeq\beta^{2}\mh\psi\simeq0.
\label{eq5}
\ee
as throughout the calculation it is assumed that $\beta\ll1$. 

From the above discussion it is clear that when $\beta$ is positive then
$\beta\ll1$ is required to 
stop the curvature perturbation from evolving on super horizon scale. But
as we have shown that $\beta\ll1$ may not be a sufficient 
condition to have the spectral index in the observed range. We have already
shown in section (\ref{ss1}) that an additional condition
$\beta<\epsilon_{V}$ is required to 
satisfy in order to have $n_{s}$ in the observed range in the branch where
$\beta$ is positive. Arbitrarily large value of $\xi$ (inconsistent with
$\beta<\epsilon_{V}$) are 
not allowed in this case.

%%%%%%%%%%%%%%%%%%%%%%%%%%%%%%%%%%%%%%%%%%%%%%%%%%%%%%%%%%%%%%%%%%%%
In conclusion we have done an explicit first order gauge invariant
perturbation theory in the Jordan frame. Unlike the work done by previous
authors where adiabatic equation of state was assumed, we have done the
calculations without assuming any form of equation of state. We show that
the results like spectral index and the bound on non-minimal coupling
constant $\xi$ is in good agreement with the known facts. For those classes
of potentials where the inflaton velocity $\dot\vp>0$ we must have a bound
on the non-minimal coupling constant $\xi$ so that the condition for the
Hubble expansion rate $0<-\frac{\dot H}{H^{2}}\ll1$ is respected. Whereas in
the other case when $\dot\vp<0$ the coupling constant $\xi$ can be very
large provided the slow-roll parameter $\epsilon_{V}$ is very small.

%%%%%%%%%%%%%%%%%%%%%%%%%%%%%%%%%%%%%%%%%%%%%%%%%%%%%%%%%%%%%%%%%%%%%%%%
\appendix
%\section{Appendix}

%%%%%%%%%%%%%%%%%%%%%%%%%%%%%%%%%%%%%%%%%%%%%%%%%%%%%%%%%%%%%%%%%%%%%%%%%%%%%%%
%\bibliographystyle{plain}
\bibliography{einsteintojordon}
\bibliographystyle{apsrev4-1}

%\end{document}
\end{document}